\documentclass[aps,showpacs,twocolumn,prd,superscriptaddress]{revtex4}

\usepackage{amsmath}
\usepackage{latexsym}
\usepackage{graphicx}

\def\Rext{R_\mathrm{ext}}

\begin{document}

\title{Binary black hole late inspiral: Simulations for gravitational wave observations}

\author{John G. Baker}
\affiliation{Gravitational Astrophysics Laboratory, NASA Goddard Space Flight Center, 8800 Greenbelt Rd., Greenbelt, MD 20771, USA}
\author{Sean T. McWilliams}
\affiliation{University of Maryland, Department of Physics, College Park, MD 20742, USA}
\author{James R. van Meter}
\affiliation{Gravitational Astrophysics Laboratory, NASA Goddard Space Flight Center, 8800 Greenbelt Rd., Greenbelt, MD 20771, USA}
\affiliation{Center for Space Science \& Technology, 
University of Maryland Baltimore
County, Physics Department, 1000 Hilltop Circle, Baltimore, MD 21250}
\author{Joan Centrella}
\affiliation{Gravitational Astrophysics Laboratory, NASA Goddard Space Flight Center, 8800 Greenbelt Rd., Greenbelt, MD 20771, USA}
\author{Dae-Il Choi}
\affiliation{Gravitational Astrophysics Laboratory, NASA Goddard Space Flight Center, 8800 Greenbelt Rd., Greenbelt, MD 20771, USA}
\affiliation{Universities Space Research Association, 10211 Wincopin Circle, Suite 500, Columbia, MD 21044}
\affiliation{Korea Institute of Science and Technology Information, \\
52-11, Eoun-Dong, Yuseong-Gu, Daejeon, South Korea, 305-806}
\author{Bernard J. Kelly}
\affiliation{Gravitational Astrophysics Laboratory, NASA Goddard Space Flight Center, 8800 Greenbelt Rd., Greenbelt, MD 20771, USA}
\author{Michael Koppitz}
\affiliation{Gravitational Astrophysics Laboratory, NASA Goddard Space Flight Center, 8800 Greenbelt Rd., Greenbelt, MD 20771, USA}
\affiliation{Albert Einstein Institute, Am M{\"u}hlenberg 1, 14471 Golm, Germany}

\date{\today}

\begin{abstract}
  Coalescing binary black hole mergers are expected to be the
  strongest gravitational wave sources for ground-based
  interferometers, such as LIGO, VIRGO, and GEO600, as well as the
  space-based interferometer LISA.  Until recently it has been
  impossible to reliably derive the predictions of General Relativity
  for the final merger stage, which takes place in the strong-field
  regime.  Recent progress in numerical relativity simulations is,
  however, revolutionizing our understanding of these systems.  We
  examine here the specific case of merging equal-mass Schwarzschild
  black holes in detail, presenting new simulations in which the black
  holes start in the late inspiral stage on orbits with very low
  eccentricity and evolve for $\sim 1200M$ through $\sim 7$ orbits
  before merging.  We study the accuracy and consistency of our
  simulations and the resulting gravitational waveforms, which
  encompass $\sim 14$ cycles before merger, and highlight the
  importance of using frequency (rather than time) to set the physical
  reference when comparing models. Matching our results to PN
  calculations for the earlier parts of the inspiral provides a
  combined waveform with less than one cycle of accumulated phase
  error through the entire coalescence.  Using this waveform, we
  calculate signal-to-noise ratios (SNRs) for iLIGO, adLIGO, and LISA,
  highlighting the contributions from the late-inspiral and
  merger-ringdown parts of the waveform, which can now be simulated
  numerically.  Contour plots of SNR as a function of $z$ and $M$ show
  that adLIGO can achieve SNR $\gtrsim 10$ for some intermediate-mass
  binary black holes (IMBBHs) out to $z \sim 1$, and that LISA can see
  massive binary black holes (MBBHs) in the range $3\times10^4
  \lesssim M/M_\odot \lesssim 10^7$ at $SNR > 100$ out to the earliest
  epochs of structure formation at $z > 15$.
\end{abstract}


\maketitle

\section{Introduction}

Coalescing binary black holes (BBHs) are strong sources of
gravitational radiation for ground-based detectors such as LIGO,
VIRGO, and GEO600, and for the space-based LISA.  This coalescence is
generally considered to proceed in three phases: inspiral, merger, and
ringdown \cite{Flanagan97a}.  During the inspiral stage, the black
holes are well separated and spiral together on quasicircular
orbits. This is followed by the merger stage, during which the black
holes plunge towards the center and merge together, forming a common
horizon.  This distorted remnant then `rings down' to form a Kerr
black hole by shedding its nonaxisymmetric modes as gravitational
radiation.

Different techniques are used to calculate the gravitational radiation
from these stages.  The inspiral can be treated analytically with
post-Newtonian (PN) techniques; the gravitational waveform is a chirp,
a sinusoid that increases in both amplitude and frequency.  The later
part of the ringdown can be handled analytically using black hole
perturbation theory, with the resulting waveforms being exponentially
damped sinusoids of constant frequency.  However, calculating the
waveforms from the dynamical merger, which produces the highest
luminosity signal, requires numerical relativity simulations of the
full Einstein equations in three spatial dimensions plus time.  With
the first generation of ground-based detectors now taking data and
LISA moving forward, knowledge of the merger waveforms and their
impact on detectability and parameter estimation is urgent.

Recently there has been dramatic progress in numerical relativity
calculations of BBH mergers.  The first full orbit of an equal mass
nonspinning BBH was achieved nearly three years ago
\cite{Bruegmann:2003aw} (see also \cite{Diener:2005mg}), using a
conformal formalism and comoving coordinates, with the black holes
represented as punctures \cite{Brandt97b}, and held fixed in the grid.
This was followed about a year and a half later by the first
simulation of the final orbit, merger, and ringdown
\cite{Pretorius:2005gq}; this work was carried out using generalized
harmonic coordinates with excised black holes moving through the
computational domain \cite{Pretorius:2004jg,Pretorius:2006tp}.
Roughly six months later the moving puncture method, which is based on
a conformal formalism and allows the puncture black holes to move
across the grid without the need for excision
\cite{Campanelli:2005dd,Baker:2005vv}, was introduced.  This simple
yet powerful technique proved to be highly effective
\cite{vanMeter:2006vi,Hannam:2006vv,Bruegmann:2006at,Sperhake:2006cy},
allowing the evolution of black holes through multiple orbits, merger,
and ringdown \cite{Campanelli:2006gf,Baker:2006yw}.  Simulations with
unequal masses \cite{Herrmann:2006ks, Baker:2006vn,Gonzalez:2006md}
and with spins \cite{Campanelli:2006uy,Campanelli:2006fg} quickly
followed.  Longer simulations with several orbits before merger have
also been carried out using excision and harmonic techniques
\cite{Buonanno:2006ui}, allowing comparisons between the resulting
waveforms \cite{Baker:comp}.

This explosion of new work on BBH mergers has opened up exciting
opportunities for applications of these results.  A number of key
applications center on gravitational wave data analysis.  In
particular, the need for accurate templates for the full gravitational
wave signal -- encompassing the inspiral, merger, and ringdown phases
-- means it is essential to understand the relationship between PN
waveforms and the results from numerical relativity
\cite{Baker:2006yw,Buonanno:2006ui,Baker:2006ha}. And, with longer
simulations now becoming available, it is becoming feasible to apply
these waveforms to questions about the detectability of BBHs for
various detectors \cite{Buonanno:2006ui}.  In this paper, we focus
especially on issues of accuracy in long simulations, starting in the
late-inspiral regime, and their applications in gravitational wave
data analysis.

The most rapid advances cover the previously least understood portion
of the radiation, the final few cycles of radiation generated from
near the ``innermost stable circular orbit'' (ISCO) and afterward,
which we call the ``merger-ringdown''.  Already there has been
considerable progress toward a full understanding of this important
portion of the waveform through which the frequency sweeps by a factor
of $\sim 3$ up to ringdown.  A significant development in this regard
was the demonstration of initial-data-independence of merger waveforms
for equal-mass, spinless black holes \cite{Baker:2006yw}.  Today,
there is general agreement that the merger of such a system produces a
final Kerr BH remnant with spin $a/m \sim 0.7$, and that the amount of
energy radiated in the form of gravitational waves, starting with the
final few orbits and proceeding through the plunge, merger and
ringdown, is $E_{\rm rad} \sim 4 \%$
\cite{Pretorius:2005gq,Campanelli:2005dd,Baker:2005vv,Campanelli:2006gf,
  Baker:2006yw}; see \cite{Centrella:2006it} for a review.  There is
also consensus on the overall simple shape of the waveforms; detailed
comparison of results among numerical relativity groups has already
begun \cite{Baker:comp}, with more to follow.  The exploration of
parameter space, including various mass ratios
\cite{Herrmann:2006ks,Baker:2006vn,Gonzalez:2006md} and spins
\cite{Campanelli:2006uy,Campanelli:2006fg}, is now underway.  Future
efforts will involve pushing this frontier to increasingly complex
mass-ratio and spin-orientation combinations, and to establishing
initial-data independence across these parameters.

Though more challenging, it is now becoming practical to simulate BBHs
starting in the late inspiral as well.  We report on new simulations
covering roughly an additional factor of 3 in frequency before the
``merger-ringdown''.  We have simulated the coalescence of equal-mass,
nonspinning black holes, starting in the late-inspiral regime, and
continuing through the merger and ringdown.  The black holes start on
orbits with very low eccentricity and spiral together through $\sim 7$
orbits before merging.  The resulting gravitational waveform has $\sim
14$ cycles, enabling a robust consistency test with late inspiral PN
waveforms as reported in \cite{Baker:2006ha}.  In
Sec.~\ref{sec:overview} we present an overview of our numerical
results.  Readers with an interest in the details of the simulations
will find them in Sec.~\ref{sec:SimAnalysis}, where we discuss our
methodology and the accuracy of our simulations, as well as introduce
the important notion of considering physical quantities as functions
of frequency rather than time when comparing predictions from
different computations.

The rapid advances in understanding merger-ringdown waveforms with
emerging simulation results for the late inspiral create a new context
for considering gravitational wave observations of this part of the
signal from such a system.  In Sec.~\ref{sec:Applic} we revisit an
analysis of merger detectability originally carried out in Flanagan
and Hughes \cite{Flanagan97a}, now in the context of presentday and
emerging waveform modeling capabilities.  In particular, we examine
the relative contributions to the signal-to-noise ratio (SNR) for the
merger-ringdown and late-inspiral portions of equal-mass coalescence
signals, as observed by initial LIGO (iLIGO), advanced LIGO (adLIGO),
or LISA.  We also plot contours of SNR as functions of redshift $z$
and total mass for adLIGO and LISA, showing their ability to detect
BBHs in the cosmos.  These SNR calculations draw on a refined waveform
model based on a best-estimate waveform using both numerical
relativity and PN results.  We conclude in Sec.~\ref{sec:summary} with
a summary and discussion of our main results.

\section{Overview of Simulation Results}
\label{sec:overview}

\begin{figure*}
\includegraphics*[scale=.53, angle=-90]{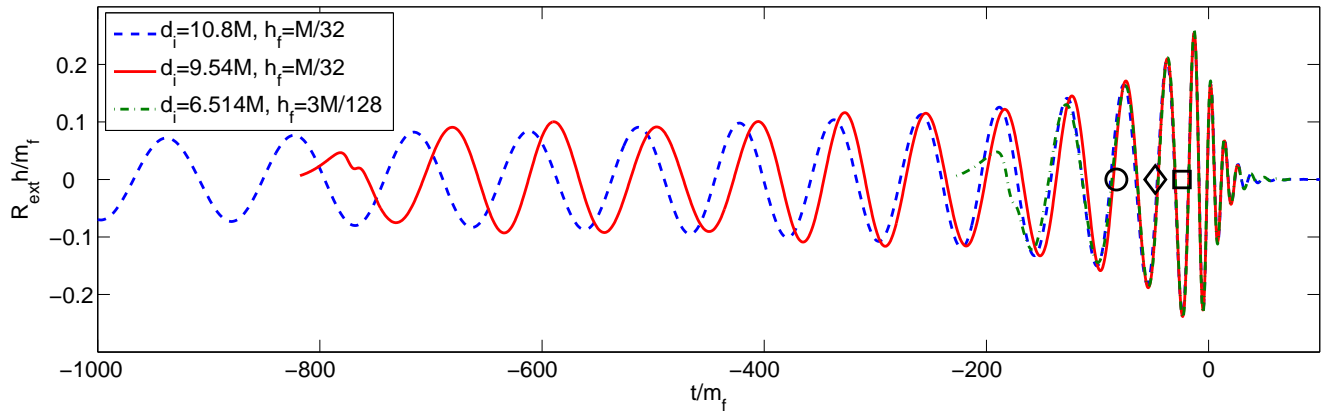}
\caption{Gravitational strain waveforms from the merger of equal-mass
  Schwarzschild black holes. The late part of the merger ($t \gtrsim
  -50M$) is robustly determined and relatively easily calculable,
  while simulations of the late inspiral (early part of the waveform)
  are rapidly approaching the phasing accuracy required for
  observational applications \cite{Baker:2006ha}.  The solid blue line
  shows our current ``best'' numerical waveform.  The dashed red line
  shows a comparison waveform from a run starting with the same
  initial data as R4 in Ref.~\cite{Baker:2006yw} and the dash-dotted
  green curve shows the results from the highest resolution R1 run in
  Ref.~\cite{Baker:2006yw}.  All waveforms have been extracted at
  $R_{\rm ext} = 40M$ and shifted in time so that the moment of
  maximum $\psi_4$ radiation amplitude occurs at time $t = 0$.  The
  initial coordinate distance between the punctures, $d_i$, is
  indicated in all cases.}
\label{fig:Wave}
\end{figure*}

Late-inspiral evolutions, covering more than a few orbits prior to
ringdown, are more challenging than shorter merger-ringdown
simulations.  In this regime there is a stronger requirement for
numerical stability and a greater demand for computational resources.
In addition, better accuracy is required to control the accumulated
phase error, which in turn constrains the numerical error that can be
tolerated in the rate of energy loss.

Using the moving puncture technique
\cite{Campanelli:2005dd,Baker:2005vv,vanMeter:2006vi}, with the gauge
evolution given by Eq.(17) and Eq.(26) in \cite{vanMeter:2006vi}, we
have simulated the evolution of a nonspinning equal-mass BBH starting
at relatively wide separation, $\sim 1200M$ or $\sim 7$ orbits before
the formation of a common event horizon.  Here $M$ is the total mass
the system would have had when the black holes were very far apart and
before radiative losses became significant.  We used
fourth-order-accurate finite differencing techniques with adaptive
mesh refinement (AMR) to resolve the dynamics near the black holes and
in the wave propagation region.  We carried out three runs using
similar grid refinement structures, but at different resolutions: low
($h_f = 3M/64$), medium ($h_f = 3M/80$) and high ($h_f = M/32$).
Here, $h_f$ is the grid spacing in the regions with the highest
resolution in each simulation, those being the regions around each
black hole.  Overall, the Hamiltonian constraint converges at fourth
order, and the momentum constraint at better than second order,
throughout the runs.

Fig.~\ref{fig:Wave} shows the resulting gravitational waveform in the
context of recent developments in black hole evolutions.  The dashed
(blue) curve gives the gravitational wave strain from the dominant
$l=2$, $m=2$ spin-weighted spherical harmonic from our highest
resolution simulation, extracted at $R_{\rm ext} = 40M$.  This
represents an observation made on the equatorial plane of the system,
where only a single polarization component contributes to the measured
strain.  Time $t=0$ is taken to be the moment of peak radiation
amplitude as measured by the Weyl curvature $\psi_4$; the symbols
along the time axis mark the points at which the system reaches the
ISCO calculated for a Schwarzschild black hole (circle), EOB 3PN
\cite{Buonanno:1998gg,Damour:2000we} (diamond), and adiabatic 3 PN
\cite{Damour:2000we,Blanchet:2001id} (square).  For comparison, we
show two other waveforms from earlier simulations carried out by this
group; both were extracted at $R_{\rm ext} = 40M$ and have been
shifted in time and initial phase so (as in \cite{Baker:2006yw}) that
the moment of peak $\psi_4$ radiation amplitude occurs at $t=0$. As
these different simulations may radiate different fractions of the
initial mass, we choose the mass $m_f$ of the \emph{post-merger} Kerr
hole as the natural length scale for comparison (see discussion of
Table \ref{table:EJ} for details). Because of radiative losses, $m_f
\approx 0.95 M$.

The solid (red) curve shows the results from a model that starts $\sim
550M$ before merger with the same initial data as run R4 in
Ref. \cite{Baker:2006yw} but using a different gauge.  Note that the
gauge conditions used for the dashed (blue) curve and the solid (red)
curve are equivalent to those given by case $\#8$ in
Ref. \cite{vanMeter:2006vi}, while the gauge conditions used in
Ref. \cite{Baker:2006yw} are equivalent to those given by case $\#3$
in Ref. \cite{vanMeter:2006vi}.  The dot-dash (green) curve shows a
simulation that starts $\sim 200M$ before merger; this is the high
resolution run R1 from Ref. \cite{Baker:2006yw}.  All three waveforms
agree to within $\sim 1\%$ for the merger-ringdown burst part of the
waveform, starting near $t \sim -50M$.

\begin{figure}
\includegraphics*[scale=.45, angle=-90]{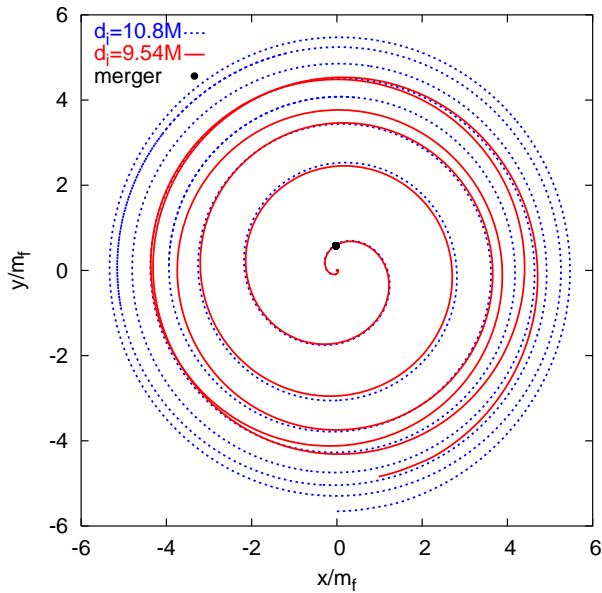}
\caption{The trajectory of one of the binary system's black holes
  through $\sim 7$ revolutions before coalescence for our high
  resolution case is shown by the dotted line.  The solid line gives
  the trajectory of the moderately long comparison run.  The initial
  coordinate distance between the punctures, $d_i$, is indicated in
  both cases.}
\label{fig:track}
\end{figure}

Astrophysically, BBHs in this relatively late stage of their evolution
are expected to be traveling on nearly circular orbits, as any initial
eccentricity would have been radiated away early in their evolution.
In this new model, we start the black holes on nearly circular orbits
with very small eccentricity $e < 0.01$, as defined below.  The
resulting black hole trajectories are shown in Fig.~\ref{fig:track},
where the tracks mark the paths of the punctures; for clarity, only a
single black hole from each simulation is shown.  The dashed (blue)
curve shows the results from our new high resolution long run; note
that the early orbits are very nearly circular.  The solid (red) curve
shows the trajectory of the moderately long comparison run (shown by
the dashed line in Fig.~\ref{fig:Wave}).  At early times, this model
is significantly more eccentric than our new results.  At later times
this trajectory locks on to that of our new run, $\sim 2.5$ orbits
before merger.

\begin{figure}
\includegraphics*[scale=.35, angle=-90]{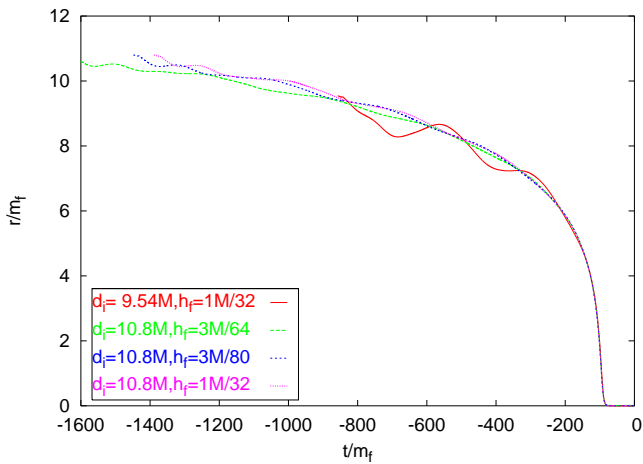}
\caption{The coordinate separation between the puncture black holes is
  shown as a function of time.  The solid line shows the results for
  the comparison run, which has relatively large eccentricity.  The
  other curves show the three resolutions for our new simulations, all
  having noticeably less eccentricity.  Note that equivalent gauge
  evolution equations were used in all four cases.}
\label{fig:rvst}
\end{figure}

The black hole separation as a function of time is shown for these two
models in Fig.~\ref{fig:rvst}.  The greater eccentricity of the
previous model (solid curve) is clearly distinguished here.  We also
show all three resolutions of our newest model.  The slight deviations
from the overall smooth trend give an indication of the small amount
of eccentricity in these latter simulations.  Note however the
differences between these three resolutions, particularly in the total
time between the start of the simulation and the merger.  Although
significant, the differences in merger time appear to converge at a
rate consistent with fourth-order error.

In the next section we consider numerical techniques and the fidelity
of the simulations in more detail, including the convergence and
conservation properties.  We also carry out a detailed analysis of the
resulting gravitational waveforms, underscoring the differences
between using time and frequency to set references for comparing
models. Readers more interested in issues of detectability and SNR
analyses for iLIGO, adLIGO, and LISA are invited to skip to
Sec.~\ref{sec:Applic}.

\section{Simulation Details}
\label{sec:SimAnalysis}

\subsection{Numerical Methodology}

For initial data we used Brandt-Br\"{u}gmann puncture
data\cite{Brandt97b}, generated by a second-order-accurate multigrid
solver, {\tt AMRMG}\cite{Brown:2004ma}.  The puncture parameters were
determined by the Tichy-Br\"{u}gmann prescription for quasicircular
initial data\cite{Tichy:2003qi}, adjusted slightly, as informed by
previous empirical experience, to reduce eccentricity.  Our adjustment
was simply to reduce the initial coordinate separation by roughly
$2\%$ while increasing the initial momentum of each puncture such that
the product of the initial momentum with the initial coordinate
separation remains constant.  The success of this approach in giving a
circular inspiral is roughly indicated by the puncture track (dotted
line) in Fig.~\ref{fig:track} and the curves in Fig.~\ref{fig:rvst}.
The puncture track was computed by integrating the equation ${\dot
  {\vec x}}_{punc}=-{\vec \beta({\vec x}_{punc})}$, where $x_{punc}$
is the position of the puncture, and ${\vec \beta}$ the shift.

This data was evolved using standard
Baumgarte-Shapiro-Shibata-Nakamura (BSSN) evolution equations, with
the addition of dissipation terms as in \cite{Huebner99} and
constraint-damping terms as in \cite{Duez:2004uh} in order to ensure
robust stability.  The dissipation terms, however, were tapered with
Gaussian functions so as to vanish at each puncture position; this
modification proved necessary for accuracy.  The gauge condition was
that recommended in \cite{vanMeter:2006vi} for moving punctures.  Time
integration was performed with a fourth-order Runge-Kutta algorithm,
and spatial derivatives with fourth-order-accurate finite-differencing
stencils.  For the outer boundary we employed a second-order-accurate
Sommerfeld condition, pushed to $|x_i|=1536M$ to keep reflections far
from the source.  AMR was implemented via the software package
PARAMESH \cite{paramesh,parameshMan}, and interpolation between
refinement regions was fifth-order-accurate.  Note that we use AMR
only to resolve the sources, and the mesh will progressively become
coarser far away from the sources.  Although the radiation which
reaches the outer boundary during the course of the simulation, with
wavelengths of $\gtrsim 100M$, will not be well resolved in this
lowest refinement region of grid-spacing $h=32M$ (in the highest
resolution run), reflections from there are causally disconnected from
our extraction radii at $R\leq 100M$.  We do not use AMR to follow the
radiation with fine mesh; instead we require only that the fixed mesh
resolution in the region of the extraction surfaces be sufficient to
resolve the waves there.  For example the extraction surface at
$R=60M$, in our highest resolution simulation, spans regions with grid
spacing $h=1M$ and $h=2M$.

\subsection{Simulation Analysis}

\begin{figure}
\includegraphics*[scale=.36, angle=-90]{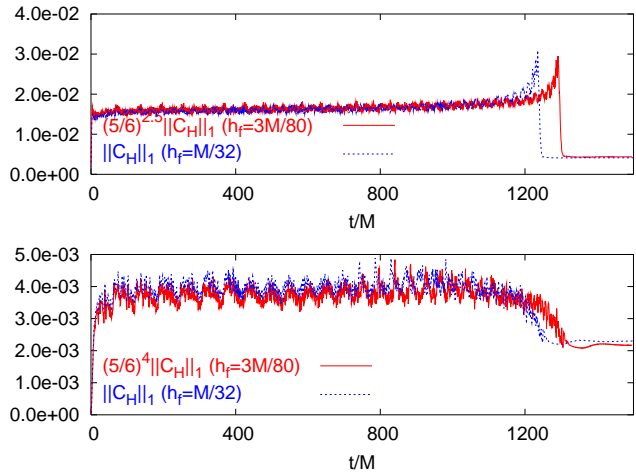}
\caption{Convergence plot for the Hamiltonian constraint $C_H$.  The
  top panel shows results from the finest grid and has been scaled so
  that for 2.5 order convergence the curves should superpose.  The
  bottom panel shows results from the second finest grid and has been
  scaled so that for fourth order convergence the curves should
  superpose; the curves indeed appear to be fourth order convergent.}
\label{fig:Ham_conv}
\end{figure}

\begin{figure}
\includegraphics*[scale=.36, angle=-90]{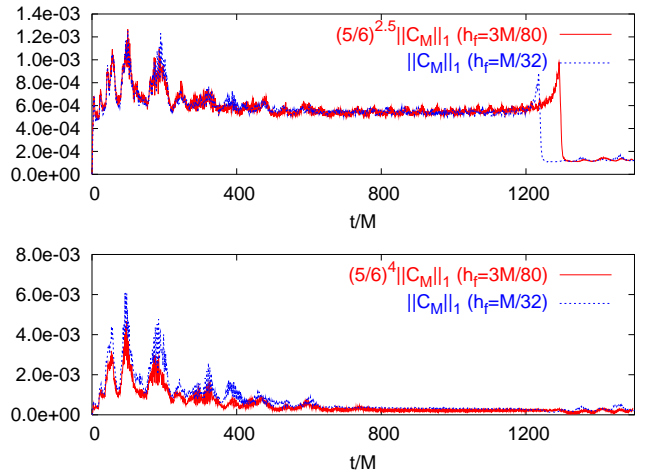}
\caption{Convergence plot for the Momentum constraint $C_M$.  The top
  panel shows results from the finest grid and has been scaled so that
  for 2.5 order convergence the curves should superpose.  The bottom
  panel shows results from the second finest grid and has been scaled
  so that for fourth-order convergence the curves should superpose;
  the curves appear less than fourth-order convergent but better than
  second-order convergent.}
\label{fig:Mom_conv}
\end{figure}

The accuracy of the simulations was assessed by various means.  We
first considered the $L_1$ norm of the constraints in each refinement
region, the grid structure having been designed to be commensurate for
all resolutions; results for the finest (top panel) and second finest
(bottom panel) refinement regions are plotted in
Fig.~\ref{fig:Ham_conv} and Fig.~\ref{fig:Mom_conv}.  The Hamiltonian
and momentum constraints were found to be convergent at an apparent
order of 2.5 in the finest grid, where the error at the puncture can
be expected to dominate and fourth-order finite differencing must
break down due to the irregularity there.  In all coarser regions the
Hamiltonian constraint appears to be fourth-order-convergent, while
the momentum constraint appears closer to second-order convergent.

\begin{figure}
\includegraphics*[scale=.36, angle=-90]{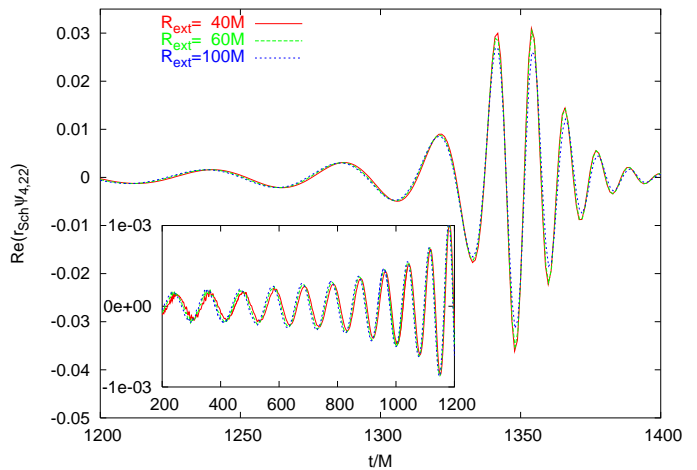}
\caption{$\psi_4$ waveform calculated at three different extraction
  radii, and scaled by the approximate Schwarzschild areal radius.
  Times have been shifted according to the approximate Schwarzschild
  ``tortoise coordinate''\cite{Misner73,Fiske:2005fx}, as appropriate
  to compare the $R_{ext}=40M$ and $R_{ext}=60M$ waveforms with the
  $R_{ext}=100M$ waveform.  }
\label{fig:psi4extraction}
\end{figure}

From each simulation we have measured the radiation in the form of the
complex Weyl tensor component $\psi_4$, specified consistently with
\cite{Baker:2001sf} to leading order in $1/r$.  The gravitational wave
strain is related to $\psi_4$ by $-\ddot{h}_+ + i \ddot{h}_\times = 2
\psi_4$, and can be recovered by integration, with the two complex
integration constants chosen to keep the strain close to oscillating
about zero. For some applications we also examine waveforms in the
form of the \emph{strain rate} $v=\dot h_+ + i \dot h_\times$, the
quantity from which radiation energy is directly obtained.  To extract
the waveform information from the simulation we define a series of
coordinate spheres of different radii $\Rext/M\in\{40,60,100\}$ on
which we measure spin-weighted spherical harmonic components of
$\psi_4$ via a second-order algorithm described in
\cite{Misner:1999ab,Fiske:2005fx}.  In this paper we focus exclusively
on the $l=2, m=2$ component of the radiation, which mirrors the $l=2,
m=-2$ component because of equatorial symmetry.  Other components are
considerably smaller, containing $\sim 1\%$ as much energy
\cite{Baker:2006yw}.

Fig.~\ref{fig:psi4extraction} compares waveforms from our
high-resolution simulation extracted on each of our three extraction
spheres with the $\Rext=40M$ and $\Rext=60M$ waveforms shifted in time
by the intervening propagation time derived based on a Schwarzschild
black hole background.  The generally good agreement for all three
waveforms indicates that potentially worrying subtleties related to
the relatively close distance of the extraction spheres, such as
nonlinear propagation effects, or tetrad-specification sensitivity, do
not seriously affect the waveforms.  On the other hand, some small
differences are evident.  For the early portion of the waveforms,
shown in the inset, the results from the closest extraction sphere
show slight differences in amplitude and phase, suggesting that
$1/\Rext^2$ radiation details are not yet insignificant here; this is
not surprising given that the extraction radius in this case is only
$\sim 1/4$ of a wavelength.  On the other hand, the waveform from the
farthest extraction radius shows signs of dissipation for the later
higher-frequency portion of the waveform.  This is because the
radiation has propagated significantly farther, through a
lower-resolution region on the computational grid, by the time it is
measured at $\Rext=100M$.  The resolution in most of the intermediate
region is $h=2M$, about six points per cycle for the ringdown
radiation.  For the rest of the paper, we primarily employ the
waveforms extracted at the intermediate distance $\Rext=60M$, which is
only weakly affected by either of the above short- and long-wavelength
effects.

\begin{table}
  \caption{\label{table:EJ}
    The parameters of the final black hole (mass and spin parameter) formed by
    merger. Here $m_r$ and $(a/m)_r$ are calculated from the complex
    ringdown frequency $\omega_r$, while $m_f$ and $(a/m)_f$ are calculated
    from the radiation measured at the given radius.  Analysis of these quantities
    at $\Rext=40M$ across the three different resolutions suggests that they
    are at least second-order convergent, and comparison across extraction radii
    gives error bars for the $h_f=M/32$ results for $m_r$, $(a/m)_r$, $m_f$ and 
    $(a/m)_f$ of $1\%$, $1\%$, $0.3\%$, and $3\%$, respectively.}
\begin{tabular}{c|c||ccc|cc}
\hline \hline
& &\multicolumn{3}{c|}{\em ringdown}&\multicolumn{2}{r}{\em conservation}\\   
$h_f$&$\Rext$&$\omega_r$       & $m_r$   & $(a/m)_r$ & $m_f$ & $(a/m)_f$ \\
\hline\hline
$3M/64$ & $40M$ &$0.551-0.083i$& 0.971 & 0.71 & 0.953 & 0.70 \\
\hline
$3M/80$ & $40M$ &$0.553-0.087i$& 0.945 & 0.68 & 0.954 & 0.72 \\
        & $60M$ &$0.553-0.088i$& 0.931 & 0.66 & 0.955 & 0.70 \\
        &$100M$ &$0.553-0.088i$& 0.931 & 0.66 & 0.959 & 0.71 \\
\hline
$ M/32$ & $40M$ &$0.553-0.085i$& 0.953 & 0.69 & 0.955 & 0.73 \\
        & $60M$ &$0.553-0.085i$& 0.953 & 0.69 & 0.955 & 0.71 \\
        &$100M$ &$0.553-0.086i$& 0.945 & 0.68 & 0.958 & 0.71 \\
\hline \hline
\end{tabular}
\end{table}
Different information in the waveforms provides independent ways to
deduce the energy and angular momentum of the final black hole
produced by the merger.  The ringdown dynamics of the black hole after
merger contains direct information about the mass and spin parameter
of the merged black hole, so that we can deduce $m_r$ and $(a/m)_r$ by
measuring the frequency and decay rate of the ringdown
radiation. Alternatively, we can measure the radiation energy
$E_{rad}$ and angular momentum content $J_{rad}$ at a given
radius. Then, since we know the initial values, conservation of energy
and angular momentum imply the values of $m_f =
M_{ADM}|_{t=0}-E_{rad}$ and $(a/m)_f =
(J_{ADM}|_{t=0}-J_{rad})/m_f^2$. Results from both methods are shown
in Table ~\ref{table:EJ}.  By comparing $m_f$ with $m_r$ and $(a/m)_r$
with $(a/m)_f$, we can verify conservation of energy and angular
momentum in the simulations.  In Table~\ref{table:EJ} we see that at
the highest resolution, energy is conserved to within $\sim1\%$ and
angular momentum to within $\sim6\%$; and the best conservation is
seen at $R_{ext}=60M$, where energy is conserved to within $\sim0.2\%$
and angular momentum to within $\sim3\%$.

To an excellent approximation, the $l=2,m=2$ harmonic of the radiation
is polarized in the form expected for radiation generated by circular
motion.  The polar representation of the $l=2$, $m=2$ component of the
complex waveform, $\psi_{4,22}=A_{\psi}(t)\exp(i \phi_\psi(t))$, is
particularly natural for circularly polarized radiation, for which
$A_{\psi}$ and $\omega=\partial\phi_{\psi}/\partial t$ vary only
slowly.  The angular polarization frequency $\omega$ then provides a
meaningful instantaneous frequency obtained directly from the
radiation, corresponding to twice the angular frequency of orbital
motion when the black holes are still separate.  Because the radiation
is measured in the weak-field region of our simulations, where gauge
dependence is minimal, this polarization frequency provides
gauge-invariant information about the binary dynamics.

If the orbital motion is eccentric, this will leave an imprint in the
radiation, causing a slight decrease in the polarization frequency of
radiation generated near apocenter. We can recognize eccentric motion
by identifying periodic deviations from a smooth monotonic trend in
the time development of the polarization frequency $\omega(t)$.

\begin{figure}
\includegraphics*[scale=.36, angle=0]{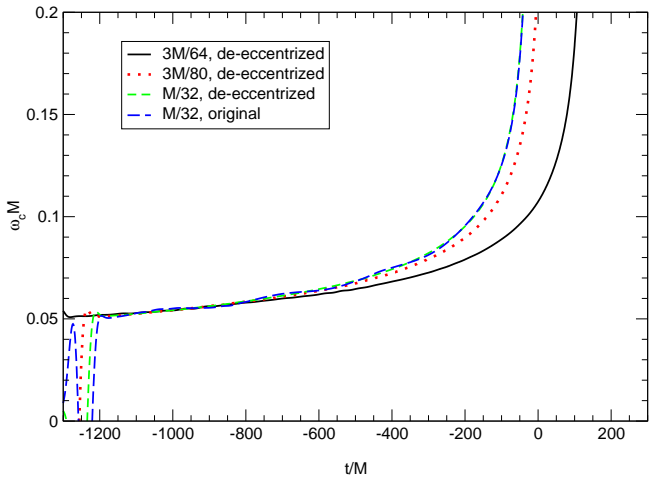}
\caption{Angular frequencies with eccentricity removed, aligned such
  that the frequencies agree when the $h_f=M/32$ simulation is $1000M$
  before the peak in $\psi_4$.  Also shown is the frequency from the
  $h_f=M/32$ resolution before the eccentricity was removed.}
\label{fig:om_c_comp}
\end{figure}

Specifically, we looked at the polarization frequency
$\omega(t)=\partial \phi/\partial t$ calculated from the strain rate
$v(t)=V(t) \exp(i\phi(t))$.  We see generally similar results whether
we use strain, strain rate, or $\psi_4$ to define the frequency, but
specifically show the strain rate because it comes out smoother than
$\psi_4$ with respect to small waveform noise, but without noticeable
detrending issues as in the strain.  We fit the time dependence of the
frequency curves to a quartic trend, $\omega_c$, plus an eccentric
modulation of the form $d\omega=A \sin(\Phi_0+\Omega_0 (t-t_0)
+\dot\Omega (t-t_0)^2)$ where the quantities $A$, $\Phi_0$, $\Omega_0$
and $\dot\Omega$ are fitting constants and $t_0=61M$ is a time offset
approximately accounting for the propagation time to $\Rext=60M$.
Ignoring the early part of the simulation where there are transient
initial data effects, and the late part where the secular trend is
very strong, we fit the curves over the time range $250 M < t <1000
M$.  We get similar results for eccentricity whether we apply
high-frequency filtering to the waveform before fitting, though we
show only low-pass filtered curves in Fig. \ref{fig:om_c_comp}. The
results of the fitting are summarized in Table~\ref{Table:ecc}.
 
\begin{table}
  \caption{\label{Table:ecc}Values resulting from eccentricity
    fitting.  The magnitude of the eccentricity in these
    simulations (as given by $AM$ or $e_0$) appears to be at least
    second-order convergent.}
\begin{tabular}{c|ccccc}
\hline \hline
Run      & $A M$       & $\Omega_0 M$ & $\dot\Omega M^2$ & $\Phi_0$ & $e_0$\\
\hline
$3M/64$  & $5\times10^{-4}$ & 0.017 & $3\times10^{-6}$ & 1   & 0.005\\ 
$3M/80$  & $7\times10^{-4}$ & 0.016 & $4\times10^{-6}$ & 1   & 0.007\\
$M/32$   & $8\times10^{-4}$ & 0.017 & $3\times10^{-6}$ & 1   & 0.008\\
\hline \hline
\end{tabular}
\end{table}
 
Variations in the details of the fitting procedure, such as using
strain instead of strain rate or smoothing high-frequency noise from
$\omega(t)$, give results consistent to roughly the number of
significant digits shown for $A$ and $\Omega_0$, though $\dot\Omega$
and $\Phi_0$ vary more significantly in some cases.  The period of
eccentric oscillations indicated by $\Omega_0$ is about 1.5 times the
initial orbital period, decreasing slightly at a rate comparable to
the rate at which the wave period grows.  Allowing $A$ to evolve in
time did not result in clearly improved fits.

From these fits we define eccentricity based on the effect, in
Keplerian dynamics, of small eccentricity on orbital frequency, which
should provide a good approximation in the adiabatic regime.  Kepler's
second law (conservation of angular momentum) implies that $L\propto
r^2\omega$ is constant, from which it follows that the eccentric
frequency deviation $d\omega/\omega$ is twice the eccentric radial
deviation $dr/r$, suggesting the unitless definition of eccentricity
$e\equiv A/2\omega_c$.  Note that, to second order in $e$, this
definition of eccentricity is also equivalent to that put forth in
Ref.\cite{Berti:2006bj} in a post-Newtonian context.  The constancy of
$A$ in our fit is consistent with a linear decrease in eccentricity as
frequency increases.  The initial eccentricity $e_0$, obtained in this
way, is also given in Table~\ref{Table:ecc}.

Even more interesting than the estimates for eccentricity provided
this way are the residual curves $\omega_c(t)=\omega(t)-d\omega(t)$ of
angular frequency vs. time with the eccentric part subtracted out.
Note that the strictly sinusoidal nature of our fit to the eccentric
modulations represented by $d\omega(t)$ implies that the underlying
secular trend should be preserved.  Fig.~\ref{fig:om_c_comp} shows the
results for each of our three simulations together with the unmodified
frequency $\omega(t)$ for the high-resolution case. The plotted curves
have also been filtered to remove some high-frequency simulation noise
present in the early part of the simulations.  The smooth, now
monotonic, trend in the curves for $\omega_c(t)$ is an indication that
most of the wiggles evident in $\omega(t)$ were consistent with our
model for eccentric modulations $d\omega(t)$, though the early part is
affected by transient features related to the shortcomings of the
initial data model.

We can take $\omega_c(t)$ as providing a record of the ``hardening''
process, as radiative losses bring the system through tightening
orbits.  The key effect of numerical simulation error evident in
Fig.~\ref{fig:om_c_comp} is a slowing of the hardening rate at low
resolutions, causing the final merger to be delayed.  This delay
appears to converge at fourth order in resolution.  Viewing the
eccentric modulation as a small perturbation on the dynamics of an
optimally noneccentric inspiral, we expect that these trends would
also provide a good approximation for the frequency evolution of a
merger simulation begun with optimally noneccentric initial data.

\subsection{Waveform accuracy}

In this section we consider the accuracy of the waveforms generated by
our simulations, focusing primarily on the accuracy of waveform
phasing information in the late-inspiral portion of our simulations.
Over the course of many developmental simulations leading up to these
results, we found that this early low-frequency part of the
simulations is the most difficult to simulate accurately. This makes
sense because the crucial dynamical details are in the slow loss of
energy and angular momentum to the relatively fine, evolving structure
of the spacetime curvature, which ultimately comprises the
radiation. Timescales are also longer for this part of the dynamics,
requiring high accuracy over a large number of computational
iterations.

\begin{figure}
\includegraphics*[scale=.36, angle=-90]{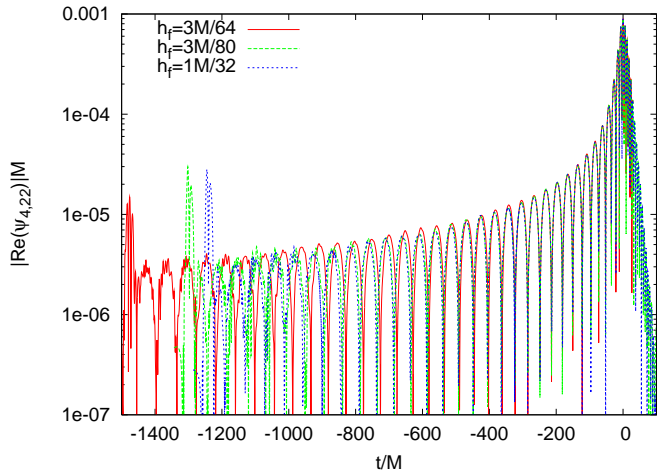}
\caption{Waveform at three resolutions.  They have been shifted in
  time and phase to agree at the peak of the wave.  The agreement
  persists backwards in time, with the growing discrepancy in phase
  converging away with resolution.}
\label{fig:wave_conv}
\end{figure}

Fig.~\ref{fig:wave_conv} compares the $\psi_4$ waveforms generated by
our simulations at different resolutions.

These waveforms have been aligned to coincide at the peak in $\psi_4$
for each of the simulations, as described in Ref.~\cite{Baker:2006yw}.
We show the waveform logarithmically because the amplitude changes by
more than two orders of magnitude through a run, part of what makes
these simulations challenging.  We are especially interested in the
phase agreement among the different simulations, evident in the nearly
vertical parts of the curves (which approach zero crossings).  Aligned
in this way, waveforms generated at all three resolutions are nearly
exactly superposed after $t\sim-250M$.  The two higher-resolution
simulations are nearly identical after $t\sim-500M$ and agree to
within a small part of a cycle for the full portion of the simulation
after the initial transient period in the first $100M$ of each run.
On the other hand, the waveforms are easily distinguished by large
differences in the overall timing of each run, with the lowest
resolution simulation going through a full additional orbit before
merger.  The high-frequency noise evident in the first part of the
simulations seems to be caused by reflections from our refinement
boundary interfaces.

\begin{figure}
\includegraphics*[scale=.36, angle=0]{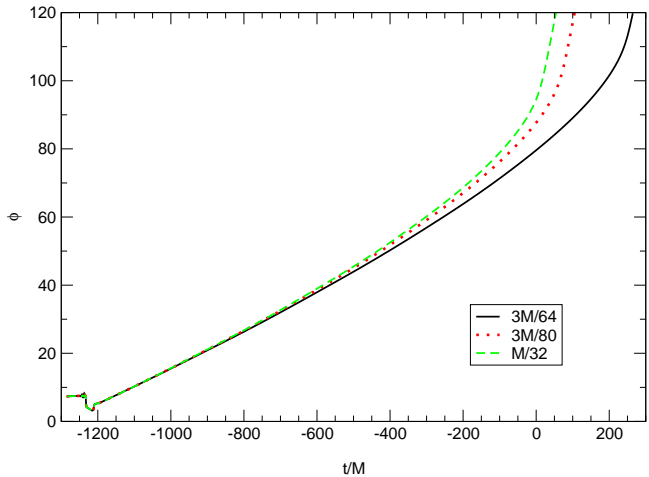}
\caption{Strain rate phase. The high-resolution simulation goes
  through about 14 cycles before merger.  The lower resolution
  simulations take longer to merge, and go through more cycles.}
\label{fig:ph_v_t}
\end{figure}

We get a more direct view of phasing information by examining the
waveforms in polar decomposition.  In Fig.~\ref{fig:ph_v_t} we show
the polarization phase derived from the strain rate waveform.

This time we have aligned the simulations in time from the beginning,
as should be appropriate for simulations of the same initial
configuration.  Later in the simulations, as the frequency increases
the phase increases more rapidly, and the timing differences among the
simulations lead to large phase differences.

\begin{figure}
\includegraphics*[scale=.36, angle=0]{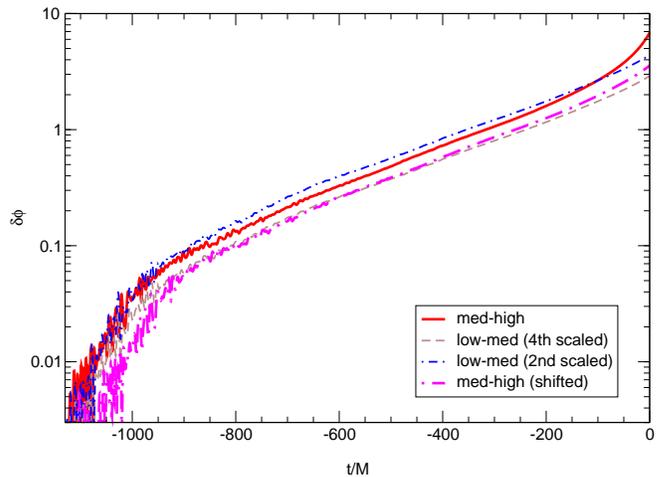}
\caption{Strain rate phase three-point convergence. The higher
  resolution phase difference is shown with and without a phase shift
  to allow comparison of errors at similar dynamical points in the
  simulation.  After shifting, the phase-error appears to be
  fourth-order convergent for much of the simulation.}
\label{fig:dph_v_t}
\end{figure}

We quantitatively compare the phasing results in
Fig.~\ref{fig:dph_v_t}, showing the difference between the two
higher-resolution runs compared with scaled versions of the difference
between the two lower resolution runs.  We can now clearly see that
the phase errors (measured this way) grow strongly in time.  The
lower-resolution difference has been rescaled two ways, such that they
would be expected to agree with the higher resolution difference for
second- and fourth-order convergence, respectively. The comparison
suggests phasing convergence somewhere between second- and
fourth-order over much of the run.

Since the phase error grows strongly in time, it is interesting to
consider the effect of timing errors in these convergence comparisons.
The last curve in the plot shows the high-medium difference shifted in
time by $57M$; that is, the high resolution and medium resolution
results have been differenced first, with their original
time-dependence unaltered, and then the resulting difference has been
shifted in time.  $57M$ represents the approximate timing difference
between the two higher resolution runs late in the simulation, as
measured from the peaks in $\psi_4$.  Viewed this way we see
time-aligned phase differences taken from similar points in the
physical evolution of the medium resolution run, as represented in the
medium-low curve, and the high resolution run, as represented in the
high-medium curve.  As a result, the runs appear to converge at a
faster rate, closer to fourth-order.  This unconventionally
time-shifted plot is not intended to serve as a rigorous assessment of
convergence, but is intended to illustrate that the timing differences
can have a significant impact on convergence estimates.

Above, we have compared our simulations made at different resolutions
by comparing the waveforms at equal points in time, with time aligned
either at the beginning of the simulation ($t=0$ in the original run)
or at the peak in $\psi_4$.  Since errors cause the simulations to
accelerate through their dynamical processes at somewhat different
rates, such comparisons end up relating moments of quite different
dynamics, and become less meaningful, depending significantly on the
reference time according to which the phases are compared. This sort
of comparison would not be applicable at all when there is no clear
way to physically align predictions in time, such as when comparing
numerical simulations with post-Newtonian (PN) results.  We can avoid
assigning a reference time by using the gravitational-wave frequency
as the reference.

For the case studied here, a quasicircular inspiral of comparable-mass
black holes, it is appropriate to consider the waveform frequency
$\omega = \partial \phi/\partial t$ as a gradually increasing
monotonic function of time, $\omega\equiv\omega(t)$.  Though the
actual waveform frequency of our simulations is not monotonic because
of small eccentricities in the inspiral trajectories, we have shown
that we can fit out the eccentric deviations. The residual secular
trend $\omega_c(t)$ provides a monotonic frequency, which we can apply
as an independent variable against which to compare various
simulations.

\begin{figure}
\includegraphics*[scale=.36, angle=0]{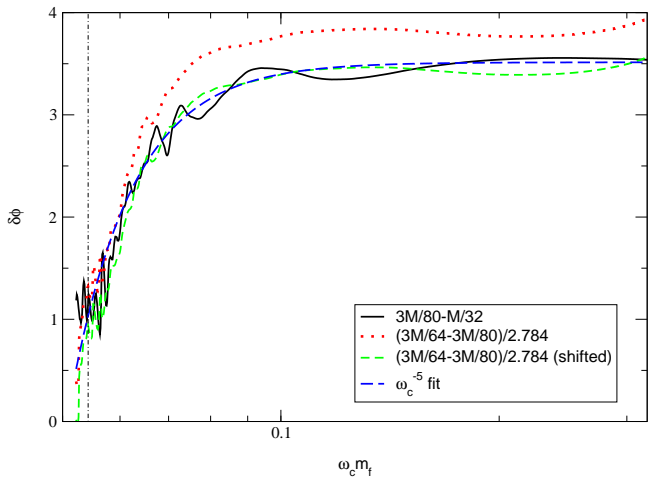}
\caption{Frequency-based phase comparisons for runs at three
  resolutions.  The solid curve represents the phase-difference
  between the $h_f=3M/80$ and $h_f=M/32$ simulations; the dotted curve
  represents the phase-difference between the $h_f=3M/64$ and
  $h_f=3M/80$ simulations; the short-dashed curve represents the
  phase-difference between the $h_f=3M/64$ and $h_f=3M/80$
  simulations, shifted vertically by a constant phase so as to agree
  with the higher resolution difference at $\omega_c m_f=0.05423$
  (shown as vertical dot-dashed line); and the long-dashed curve shows
  a fit to $\omega^{-5}$ phasing error, which might be expected if
  there are energy conservation violations that are constant in time.
  The lower-resolution curves have been scaled so that they should
  superpose with the higher-resolution curves in the case of
  fourth-order convergence.  The unshifted phase-difference curves
  appear better than fourth-order convergent, while shifting the
  lower-resolution curve makes the rate of convergence look closer to
  fourth-order.}
\label{fig:dph_v_om}
\end{figure}

We show frequency-based phase differences among the simulations in
Fig.~\ref{fig:dph_v_om}, which allows us to compare the difference
between the two higher-resolution simulations with that between the
two lower-resolution simulations.  If the simulation errors are
fourth-order-convergent, then the low-medium difference should be
approximately $2.784$ times the medium-high difference. As is evident
from the medium-high ``(3M/80-M/32)'' and low-medium
``(3M/64-3M/80)/2.784'' curves in Fig.~\ref{fig:dph_v_om}, the errors
appear slightly over-convergent with respect to fourth-order scaling.
This may be caused by phase error accrued early in the
lowest-resolution ($h_f=3M/64$) simulation, due to difficulty in
resolving high-frequency components in the spurious Bowen-York
radiation, as well as in an initial gauge pulse, which dominate at
this time.  This lowest resolution may therefore not quite be in the
convergent regime during this early part of the simulation.  If the
phases are adjusted by a constant such that they match at some point
after the main part of the Bowen-York pulse has passed, then
fourth-order scaling fits more closely.  This is demonstrated in
Fig.~\ref{fig:dph_v_om} by the ``(3M/64-3M/80)/2.784 (shifted)''
curve, which has been vertically phase-shifted by a constant so as to
agree with the ``(3M/80-M/32)'' curve at $\omega m_f=0.05423$, the
frequency $1000M$ before merger in our high-resolution simulation.

As is also clear in the figure, phase error accumulates most
significantly in the low-frequency portion of the simulation, $m_f
\omega\lesssim0.08$.  This makes sense generally since the simulation
spends much more time at lower frequencies.  Consider, for instance,
the effect of a small nonconservative leakage of energy $\delta \dot
E$ from the simulation.  When the black holes are well separated, the
dynamical development manifested in the sweeping frequency is driven
by a slow loss of energy and angular momentum.  An energy leakage
$\delta \dot E$ will change the frequency sweep-rate $\dot\omega$ by
$\delta\dot\omega\sim(\delta\dot E)/(\partial E/\partial\omega)$,
where $\partial E/\partial\omega$ indicates how the binding energy
changes with frequency.  Phase error $\delta\phi$ can be determined
from the error in the sweep-rate by
\begin{equation}
\delta\phi = \delta\int{\frac{\omega}{\dot\omega}\,\mathrm{d}\omega} = - \int{\frac{\omega}{\dot\omega^2}\delta\dot\omega\,\mathrm{d}\omega}
\end{equation}
 
Applying the leading-PN-order expressions for $\partial
E/\partial\omega$ and $\dot\omega(\omega)$ gives a result proportional
to $\omega^{-5}\delta\dot E$.  Indeed this dependence fits our phase
differences rather well, as shown in Fig.~\ref{fig:dph_v_om}. Note,
however, that this does not single out energy nonconservation as the
source of error, as other errors may produce similar effects.  A
similar leakage of angular momentum would lead to error proportional
to $\omega^{-4}$, which also fits reasonably well.

The convergence evidenced by Fig.~\ref{fig:dph_v_om} suggests that we
can apply Richardson extrapolation to estimate the difference of our
high-resolution ($M/32$) run from the infinite-resolution limit.  If
we assumed the fourth-order convergence suggested by
Fig.~\ref{fig:dph_v_om}, the phase error estimate for this run would
be $0.93$ times the difference between the phases from the $3M/80$ and
$M/32$ resolution runs.  However we will simply take the more
conservative estimate of the actual difference between these
resolutions.  Note that during the last $1000M$ of our $M/32$
simulation (i.e. from $\omega m_f=0.05423$ onwards), we estimate
roughly two and a half radians of phase error accumulate, as measured
with respect to frequency, which is less than half of a gravitational
wave cycle.  A benchmark for accumulated waveform phasing errors is
one-half of a cycle, because phase error exceeding this amount would
lead to destructive interference in matched-filtering applications.
For our high-resolution simulation, we estimate less than one-half
cycle of gravitational wave phase error over the full simulated
waveform, excluding the meaningless transients in the first $100M$.
As in \cite{Baker:2006ha}, we estimate that these phasing errors are
smaller than the implicit phasing difference between the 3PN and 3.5PN
expansions of $\dot\omega(\omega)$ after $t\sim-300M$ ($\omega_c
M\sim0.08$).  For our data analysis considerations we will only be
using the numerical waveform after this point, for which the estimated
phase error is well below a half cycle.

Frequency-based phase comparisons, such as we have presented here, are
better suited than time-based phase differences, which depend strongly
on where the waveforms are chosen to be aligned in time.  The
relationship between the two can be understood by considering a
one-parameter family of waveform results, with a parameter $\lambda$
representing model dependence, in this case the numerical
grid-spacing. The waveforms would provide phase as a function of time
$\phi_\lambda(t)$, from which we can derive frequency
$\omega_\lambda(t)$, which is monotonic for small eccentricity.
Inverting to obtain $t_\lambda(\omega)$ one can derive the
frequency-based phasing
$\bar\phi_\lambda(\omega)\equiv\phi_\lambda(t_\lambda(\omega))$.  Now
considering variations $\delta\equiv d/d\lambda$ near $\lambda=0$, one
finds the relationship between frequency- and time-based phase
comparisons
\begin{equation}
\delta\bar\phi(\omega)=\delta\phi(t(\omega))+
\omega\delta t(\omega).  
\end{equation}
This sheds some light on the often confusing issue of time alignment
in the time-based comparisons shown earlier.  Specifically, for a
waveform that sweeps significantly through frequency, time- and
frequency-based phase differences will be most similar when the
time-based phase differences are aligned so that $\delta t$ vanishes
where $\omega$ is largest.  In our case, the net frequency-based phase
differences in Fig.~\ref{fig:dph_v_om} are closer to net phase
differences with time aligned at the end of the waveform, as in
Fig.~\ref{fig:wave_conv}, than when time is aligned at the
lower-frequency beginning, as in
Figs.~\ref{fig:ph_v_t}~-~\ref{fig:dph_v_t}.

\section{applications to gravitational wave observations of BBHs}
\label{sec:Applic}

\begin{figure*}
\includegraphics*[scale=.53, angle=0]{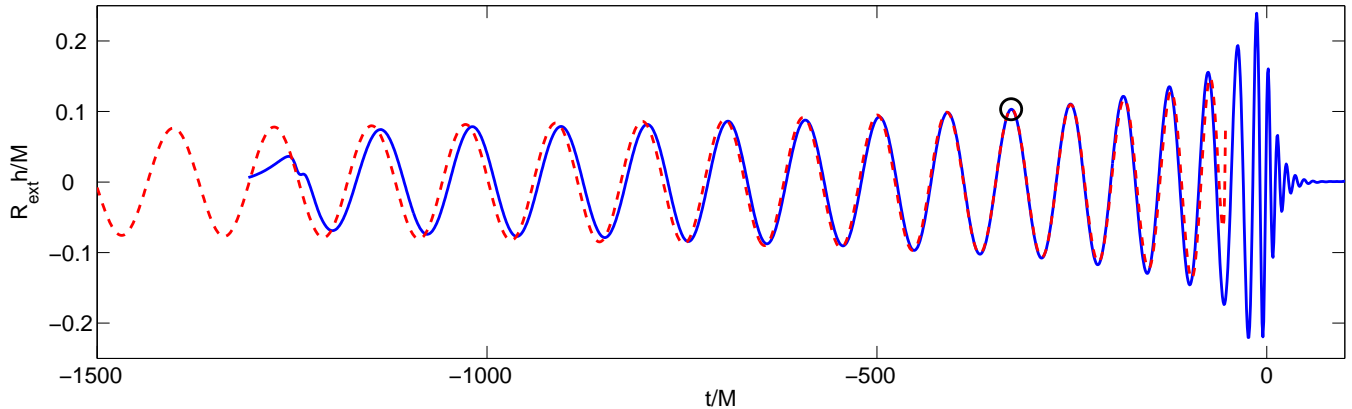}
\caption{Numerical simulation results for gravitational wave strain
  compared with post-Newtonian estimates.  The waveform shown is from
  the high resolution numerical simulation presented in
  Sec.~\ref{sec:overview} overlaid here with a PN waveform with 3.5PN
  order phasing and 2.5PN order amplitude accuracy.  The combined
  waveform, joined at $t=-328M$ (circle) is applied in
  Sec.~\ref{sec:Applic} to calculate signal-to-noise ratios for iLIGO,
  adLIGO, and LISA.}
\label{fig:Wavemain}
\end{figure*}

In the first part of this paper, we presented a state-of-the-art
calculation of the late inspiral and merger of equal mass, nonspinning
BBHs, starting $\sim 7$ orbits before the peak radiation amplitude.
In Ref.~\cite{Baker:2006ha} we showed that there is a significant
region over which the waveform from the $M/32$ numerical simulation
presented above agrees with waveforms derived from PN calculations.
In this section we will use the best of both treatments by joining the
final segment of the BBH evolution with the PN approximation for the
preceding inspiral, as shown in Fig.~\ref{fig:Wavemain}.  We will show
below that the phasing information for this waveform has an estimated
accuracy of better than one-half of a gravitational wave cycle, making
the waveform applicable in a variety of matched-filtering-based
gravitational wave data analysis applications.  We use this waveform
to calculate the relative detectability of the inspiral and the
merger-ringdown, as well as the detectability of the entire waveform,
for iLIGO, adLIGO, and LISA.  We show examples of characteristic
signal strains for both classes of detectors.  We also compute SNRs
for astrophysically interesting BBH masses, highlighting the
importance of the late inspiral and merger parts of the signal.
Modifications of our results that may arise from BBHs with spins and
unequal masses are considered.

\subsection{Waveform matching and SNR calculation}

The PN waveform used for all the analyses in this section is of order
3.5PN in phase (as derived in \cite{Blanchet:2001ax,Blanchet:2004ek})
and 2.5PN beyond the quadrupole approximation in amplitude (as derived
in \cite{Arun:2004ff}).  It should be noted that the PN approximation
that we use for the phase is actually an expansion of the chirp rate
${\dot\omega}$ in terms of the frequency $\omega$, which we then
numerically integrate.  Direct numerical integration of the 3.5PN
expansion of the chirp rate, ${\dot\omega}_{3.5PN}$, for example in
the integrand $d\omega(\omega/{\dot\omega}_{3.5PN})$, does not
strictly respect the PN approximation of the phase, as the latter
would require additional 3.5PN expansion of the integrand
itself. However, the phase obtained in this manner will have the same
convergence properties as the original ${\dot\omega}_{3.5PN}$
expansion, which is arguably a more fundamental PN quantity because of
its close relationship to the rate of energy loss, ${\dot E}$, from
which it is derived.  Additional expansion of the phase appears to
compromise its accuracy.  Using shorter runs it had been observed in
\cite{Buonanno:2006ui} that the phase obtained from numerical
integration of the PN expansion of the chirp rate seemed to agree
better with numerical simulations than did the analytic expressions;
more recently it was demonstrated in \cite{Baker:2006ha}, using runs
extending well into the late inspiral and with the effects of
eccentricity mitigated, that the numerically integrated phase appears
to be converging to the numerical result at frequencies where the full
3.5PN expansion of the phase is clearly invalid.  We therefore use the
numerically integrated 3.5PN chirp rate for the phasing of the PN
portion of our waveform.
 
Fig.~\ref{fig:Wavemain} shows our numerical waveform from Sec
\ref{sec:overview} overlaid with the PN waveform that was just
described.  To generate a complete, mass-scalable waveform, we match
the frequency of the numerical simulation to the PN prediction,
adjusting the phases to also be equal at that point as shown in
Fig.~\ref{fig:Wavemain}, and connecting the two halves to make a
single waveform.  This is done by shifting the PN waveform until the
frequency equals the numerically-predicted frequency at a time in the
simulation where the accuracy of the numerical data first surpasses
the accuracy of the PN approximation, as estimated in
\cite{Baker:2006ha}.  Specifically, \cite{Baker:2006ha} predicts this
point of equivalent accuracy to occur at $M\omega \sim 0.08$, which
corresponds to $t=-328M$ (shown by the circle in
Fig.~\ref{fig:Wavemain}).  It is worth noting that there was no need
to adjust the PN amplitude for continuity.  The amplitude agreement
with the numerical simulation is so good, and hence the resulting
amplitude is so nearly continuous, that the small discontinuity fails
to produce any discernible artifacts in the Fourier transform
$\tilde{h}(f)$ of the resulting waveform.

Having generated a waveform, it is informative to estimate the
waveform's phasing accuracy over the course of the BBH evolution.
Note in Fig.~\ref{fig:dph_v_om} that for the portion of our $M/32$
simulation that is used in the waveform, we estimate $\sim 0.5$
radians of phase error.  If we take the difference between 3 and 3.5
PN terms to be an estimate of the phase error as in
\cite{Baker:2006ha}, we can assess the error for the PN portion of the
waveform.  It was shown in \cite{McWilliams} that the analytic PN
phase expression accumulates very little error, on the order of $0.1$
radians, until $M\omega \sim 1 \times 10^{-4}$.  Beginning our
numerical phase integration at this point and evaluating up to
$M\omega = 0.08$ yields a gravitational wave phase error of $\sim 3.6$
radians, such that the total accumulated phase error over the entire
waveform is $\sim 4$ radians.  As stated previously, an accumulated
waveform phasing error of less than $\pi$ radians is the threshold
below which wave-matching comparisons may be used for
matched-filtering applications. We estimate that our combined waveform
meets this criterion after a frequency of about $M\omega\sim0.01$ up
to the ringdown frequency, $M\omega\sim0.5$.  We therefore have a
waveform with sufficient accuracy to be useful as a template for
gravitational wave detection.  While templates will ultimately be
needed for cases of greater astrophysical interest, and still greater
accuracy will be required for the template to be useful for the
purpose of parameter estimation, the construction of this waveform
illustrates that the field of numerical relativity has matured to the
point of being capable of producing results that are useful for
gravitational wave data analysis.

The calculated waveform that we have just described is actually the
total strain on the equatorial plane, where $h_{\times}$ vanishes and
therefore $h_{+}$ provides the only contribution.  To get the
optimally-oriented strain amplitude (which is the total strain passing
an observer on the equatorial axis), we multiply this result by $2
\sqrt{2}$, which is the ratio of peak total gravitational wave
amplitude to the amplitude of $h_{+}$ alone in the quadrupole
approximation.  We then simply divide by $\sqrt{5}$ in order to
convert to an orientation-averaged waveform for our subsequent
analyses.  This factor can be understood by observing that
orientation-averaging is fully equivalent to averaging over all sky
positions of the detector from the perspective of the BBH, and such
sky-averaging results in a factor $1/\sqrt{5}$ in sensitivity
\cite{Flanagan97a}.

The SNR is calculated assuming matched-filtering is performed on the
data, and that the waveforms are perfect copies of the embedded
signal.  In this case, the sky- and waveform-polarization-averaged SNR
is given by
\begin{equation}\label{eq:snr}
  <(SNR)^2> = \int d(\ln f)\left(\frac{h_{char}(f)}{h_{n}(f)}\right)^2,
\end{equation}
where $h_{char}(f)\equiv 2f\left|\tilde{h}(f)\right|$ is the
characteristic signal strain and $h_{n}(f)\equiv \sqrt{5}h_{rms}(f) =
\sqrt{5fS_n(f)}$ is the rms of the detector noise fluctuations
multiplied by $\sqrt{5}$ for sky-averaging, with $\tilde{h}(f)$ and
$S_n(f)$ being the Fourier transform of the signal strain and the
power spectral density of the detector noise, respectively
\cite{Flanagan97a}.

The waveform scales with luminosity distance $D_L$ and total mass $M$
as $h_{char}\propto(1+z)M/D_L$, while the time axis for an observed
wave, after redshifting, scales as $t\propto(1+z)M$, so that the
waveform shown in Fig.~\ref{fig:Wavemain} is applicable over all total
masses and redshifts.  When needed we relate luminosity distance to
redshift $z$ using cosmological parameters consistent with the most
recent Wilkinson Microwave Anisotropy Probe (WMAP) results
($\Omega_{\Lambda}=1-\Omega_M=0.72,\,h=0.73$) \cite{WMAP} and the
relation
\begin{equation}\label{eq:dlz}
D_L(z) = \frac{(1+z)c}{H_0}\int_0^z \frac{dz^\prime}{\sqrt{\Omega_M(1+z')^3 + \Omega_\Lambda}}.
\end{equation} 
 
For cases where an impractically long time series would be needed to
cover the band with an adequate sampling rate, the waveform is
extended in Fourier space to still lower frequencies (and consequently
back further in time) using the quadrupole formula,
\begin{equation}\label{eq:hquad}
|\tilde{h}_{{\rm quad}}(f)|=\frac{1}{2\sqrt{15}D_L}\left(\frac{[(1+z)M]^5}{\pi^4f^7}\right)^{1/6}.
\end{equation}
 
The PN portion of the waveform continues slightly past where its
Fourier transform deviates from the quadrupole expression for
$|\tilde{h}(f)|$ by $\sim 2\%$, at which point (\ref{eq:hquad}) is
used to extend $|\tilde{h}(f)|$ back as far as necessary.  The PN
segment is truncated at a higher frequency than the lowest frequency
component of its Fourier transform in order to eliminate edge effects.
Finally, the quasinormal ringdown at the end of the numerical
simulation is extended by fitting a damping coefficient and
fundamental frequency to the data in order to mitigate edge effects at
the high-frequency end of the Fourier transform.

\subsection{Observing Stellar BBHs and IMBBHs with iLIGO and adLIGO}

Ground-based interferometers are sensitive to relatively high
frequency gravitational waves from coalescing stellar mass ($M
\lesssim 10^2 M_{\odot}$) and intermediate mass (IM) ($10^2 M_{\odot}
\lesssim M \lesssim 10^3 M_{\odot}$) BBHs.  In this section, we apply
our combined waveform to consider the response of iLIGO and adLIGO to
BBH coalescence, illustrating the importance of numerical simulation
results for ground-based detectors.  LIGO has facilities in Hanford,
WA and Livingston, LA; each facility has an interferometer consisting
of two 4 km long arms, and Hanford also has a 2 km detector.  Both the
iLIGO and adLIGO detectors are designed to detect high-frequency
gravitational waves, with iLIGO sensitive in the frequency range
$40{\rm Hz} \lesssim f \lesssim 8000{\rm Hz}$ and adLIGO in the range
$14{\rm Hz} \lesssim f \lesssim 10^3{\rm Hz}$. Initial LIGO is
currently operating at or near the initial design sensitivity in a
year-long scientific data-taking run.  Advanced LIGO is a planned
upgrade that will increase the detector sensitivity by roughly an
order of magnitude across the band.  In addition, adLIGO can be tuned
to optimize its sensitivity for different sources.

For our analysis of iLIGO, we used the design sensitivity to
characterize the detector noise \cite{LIGOSRD}.  This sensitivity
assumes that the noise is seismically limited below 40 Hz, thermally
limited between 40 and 150 Hz, and shot-limited above 150 Hz.  For
adLIGO, unlike iLIGO, we had a choice of tuning configurations.  We
used the wide-band tuning typically associated with burst sources
because of its dramatically superior sensitivity at higher
frequencies, where the merger portion from many sources is predicted
to occur \cite{DHSCom}.  This yielded an improved SNR for most masses
compared to tunings that were optimized for only the early inspiral
portion of the coalescence.

\begin{figure}
\includegraphics*[scale=.21, angle=0]{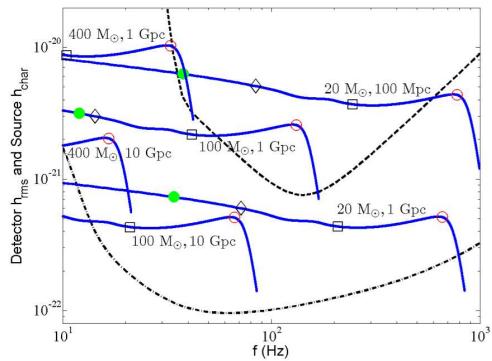}
\caption{The iLIGO (dashed) and adLIGO (dash-dotted) rms noise
  amplitudes $h_{n}$ with the characteristic amplitudes $h_{char}$ of
  6 example sources (solid).  The locations on each $h_{char}$
  corresponding to the peak $\psi_4$ amplitude (circle) and 1 second
  before the peak in the observer's frame (filled circle), as well as
  $t=-50M$ (square) and $t=-1000M$ (diamond) in the source's frame,
  are as marked.  The mass given is the combined rest mass of each
  black hole.}
\label{fig:LIGOhchar}
\end{figure}

In Fig.~\ref{fig:LIGOhchar}, we show $h_{char}$ for several sources
plotted relative to the $h_{rms}$ sensitivity curves for iLIGO (dashed
line) and adLIGO (dash-dotted line). We plot these values because the
height of $h_{char}$ above $h_{rms}$ is an indicator of the SNR, as
can be seen by inspecting Eq.~\ref{eq:snr}.

\begin{figure}
\includegraphics*[scale=.28, angle=0]{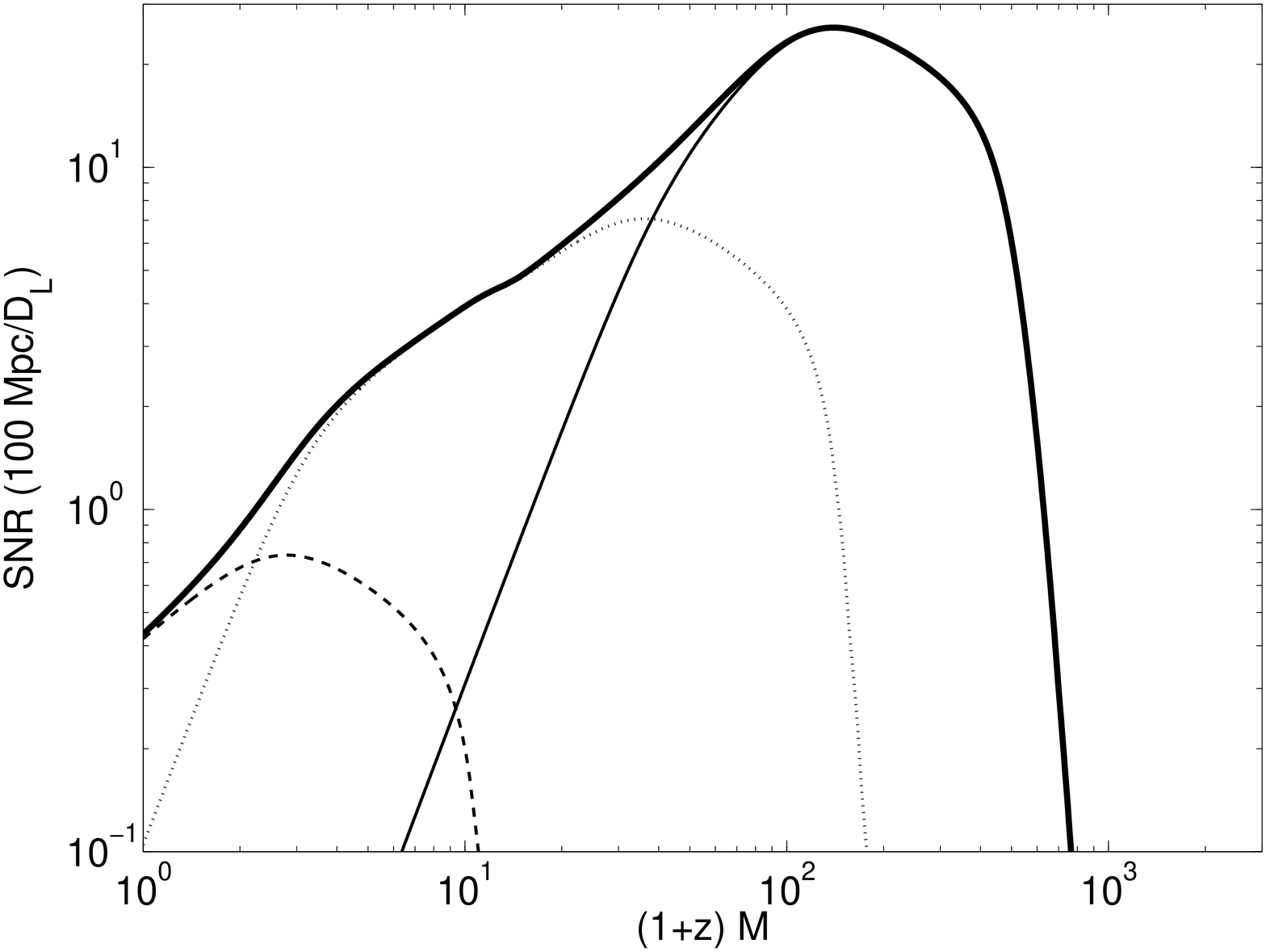}
\caption{SNR for sources at luminosity distance $D_L = 100$ Mpc
  plotted vs. redshifted mass for iLIGO.  The contributions from
  $-\infty < t < -1000M$ (dashed), $-1000 < t < -50M$ (dotted), and
  $-50M < t < \infty$ (thinner solid), as well as the SNR from the
  entire waveform (thicker solid) are shown.}
\label{fig:LIGOSNR}
\end{figure}

By rescaling we can calculate the SNR as a function of redshifted
mass, and particular luminosity distance $D_L$.  In
Fig.~\ref{fig:LIGOSNR} we plot the SNR achievable by iLIGO for sources
at a luminosity distance $D_L = 100$ Mpc as a function of redshifted
mass $(1+z)M$.  Here, the dashed line shows the SNR from the early
inspiral in the time range $- \infty < t < -1000M$, which is roughly
up to the start of our run. The dotted line shows the SNR for the late
inspiral, $-1000M < t < -50M$, where $t = -50M$ is approximately the
time at which the merger burst begins.  The thin solid line gives the
SNR for $-50M < t < \infty$, and encompasses the merger-ringdown part
of the signal.  The thick solid line shows the SNR from the entire
waveform.  Note that the addition of the merger-ringdown waveforms
increases the SNR and extends the detectable mass range significantly.
The merger-ringdown portion $t>-50M$ dominates for all equal-mass
nonspinning merger observations detectable with SNR larger than 10 at
100 Mpc.
 
This type of plot was first made in Ref. \cite{Flanagan97a}, and it is
useful to compare our results with theirs. Our SNR calculations are
based on a full waveform for the case of equal-mass, nonspinning black
holes.  The work in Ref. \cite{Flanagan97a} was done before merger
waveforms were calculated and thus is based on estimates for the
merger-ringdown regime.  For example, they estimated a merger
radiation efficiency of $\sim 10\%$, which is higher than our results
but may well obtain for mergers with spin.  Comparing their Fig. 4 for
the SNR for iLIGO with our Fig.~\ref{fig:LIGOSNR} we note that their
curve for the inspiral includes the radiation up to the merger and so
should be compared to the combination (in quadrature) of our dashed
and dotted curves.  Our result for the merger SNR is somewhat smaller
than theirs, due to the smaller amount of radiation emitted in our
mergers.  More recently, an analysis of SNR for iLIGO using numerical
relativity waveforms for the merger and PN waveforms for the inspiral
was made in Ref. \cite{Buonanno:2006ui}; our results in
Fig.~\ref{fig:LIGOSNR} are similar to what they report in their
Fig. 22.

\begin{figure}
\includegraphics*[scale=.28, angle=0]{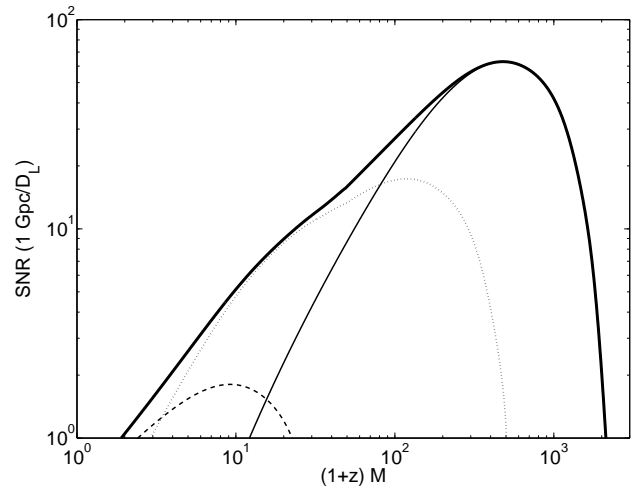}
\caption{SNR for sources at luminosity distance $D_L = 1$ Gpc plotted
  vs. redshifted mass for adLIGO.  The contributions from $-\infty < t
  < -1000M$ (dashed), $-1000 < t < -50M$ (dotted), and $-50M < t <
  \infty$ (thinner solid), as well as the SNR from the entire waveform
  (thicker solid) are shown.}
\label{fig:AdvLIGOSNR}
\end{figure}

\begin{figure}
\includegraphics*[scale=.28, angle=0]{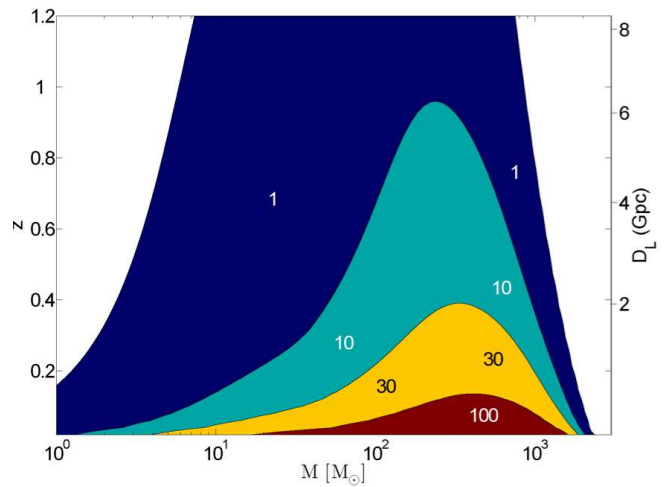}
\caption{SNR contour plot with mass and redshift dependence for adLIGO.}
\label{fig:AdvLIGOSNRcontour}
\end{figure}

Fig.~\ref{fig:AdvLIGOSNR} shows the SNR for sources at $D_L = 1$ Gpc
for adLIGO.  Comparing with Fig.~\ref{fig:LIGOSNR}, we see that adLIGO
will have a significantly higher sensitivity to BBHs over iLIGO.  This
point is reinforced in Fig.~\ref{fig:AdvLIGOSNRcontour}, which shows
contours of SNR for adLIGO as functions of redshift $z$ and total mass
$M$.  We find that for $M\sim 200M_{\odot}$, adLIGO should be able to
achieve an SNR greater than 10 out to nearly $z=1$ for equal-mass
nonspinning binaries.  From Fig.~\ref{fig:LIGOSNR} it is evident that
these high SNRs depend strongly on the merger-ringdown part of the
waveform $t>-50M$.
 
It is important to note that astrophysical BBHs are likely to have
mass ratios different from unity, and that this will reduce the SNRs
computed here for the equal-mass case.  For stellar BBHs, current work
\cite{Belczynski:2006zi} shows that the mass ratios are rather broadly
distributed.  The rates for such mergers may be low, $\sim 2 yr^{-1}$
for adLIGO, depending on the evolution of the original binary through
the common envelope phase.  For IMBBHs, mass ratios in the range $0.1
\lesssim m_1/m_2 \lesssim 1$ are expected to be the most relevant,
with potential rates of $\sim 10$ per year \cite{Fregeau:2006yz},
although these rates are far more uncertain than those for stellar
BBHs.  We can apply the mass scalings from Ref. \cite{Flanagan97a} to
show the effect of mass ratios on the computed SNRs; specifically, SNR
$\sim \eta^{1/2}$ for the inspiral, and SNR $\sim \eta$ for the merger
and ringdown, where $\eta = \mu/M$ and $\mu = m_1m_2/M$ is the reduced
mass.  Astrophysical BBHs are also expected to be spinning and this
can potentially raise the SNR, for example if there is a spin-induced
hangup that generates more gravitational wave cycles in the merger
\cite{Campanelli:2006uy}.

\subsection{Observing MBBHs with LISA}

LISA, a proposed space-based interferometer consisting of three 5
million km long arms, will be sensitive to low-frequency gravitational
waves from coalescing MBBHs in the band $3\times 10^{-5}{\rm Hz}
\lesssim f \lesssim 1{\rm Hz}$.  The coalescing massive binary black
holes (MBBHs) that radiate in this band will have masses $M \gtrsim
10^4 M_{\odot}$.

\begin{figure}
\includegraphics*[scale=.21, angle=0]{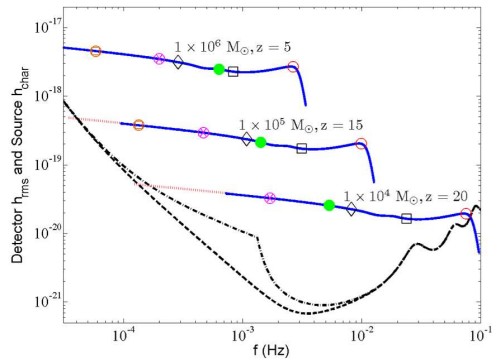}
\caption{LISA rms noise amplitude $h_{rms}$ from the detector only
  (dashed) and from the detector combined with the anticipated white
  dwarf binary confusion (dash-dotted) \cite{Barack:2004} with the
  characteristic amplitudes $h_{char}$ of three example sources
  (solid).  The locations on each $h_{char}$ curve corresponding to
  the peak $\psi_4$ amplitude (circle), 1 hour before the peak (filled
  circle), 1 day before the peak (circle with inscribed cross), and 1
  month before the peak (circle with inscribed square) in the
  observer's frame, as well as $t=-50M$ (square) and $t=-1000M$
  (diamond) in the source's frame, are as marked.  The mass given is
  the combined rest mass of each black hole.  When necessary, the
  quadrupole approximation is used to extend $h_{char}$ backward in
  time 3 years before the peak $\psi_4$ amplitude in the detector's
  frame (dotted).}
\label{fig:LISAhchar}
\end{figure}

Fig.~\ref{fig:LISAhchar} shows $h_{char}$ for several MBBHs plotted
relative to the LISA sensitivity curve.  We used the ``standard'' LISA
sensitivity curve \cite{Larson,LISASenGen} for frequencies above
$1\times 10^{-4}$ Hz, with shot and pointing noise contributions
totaling $20 {\rm pm}/\sqrt{{\rm Hz}}$ of laser phase noise.  For
$3\times 10^{-5}{\rm Hz} \leq f \leq 1\times 10^{-4}{\rm Hz}$, we
employed a more conservative estimate of the acceleration noise than
the one given in \cite{Larson}, instead assuming a steeper amplitude
spectral density that falls off as $f^{-3}$ constrained to match the
standard sensitivity curve at $1\times 10^{-4}{\rm Hz}$
\cite{MerkowitzCom}.  Below $3\times 10^{-5}{\rm Hz}$, we assume the
detector has no sensitivity, which is a reflection of the uncertainty
of the sensitivity at such low frequencies and our desire to make
conservative estimates.  The sensitivity model assumes that there are
no correlated noise sources, and its power characterization
corresponds to a round trip through one arm of the interferometer.
 
MBBH sources can remain in-band for LISA over a very broad frequency
range.  Therefore, unlike the case of iLIGO and adLIGO, LISA sources
nearly always require the use of the quadrupole approximation
procedure mentioned above to extend $h_{char}$ to sufficiently low
frequencies.  Also, since more massive BBHs chirp more slowly, a MBBH
could potentially be in LISA's sensitive band for much longer than the
mission's lifetime.  To prevent unrealistic SNR values due to this
excessive integration time, the quadrupole formula is only used to
extend $h_{char}$ to a low enough frequency such that the total
$h_{char}$ used in our calculations corresponds to 3 years of data in
the detector's frame, which is a conservative estimate of the expected
mission lifetime.

\begin{figure}
\includegraphics*[scale=.28, angle=0]{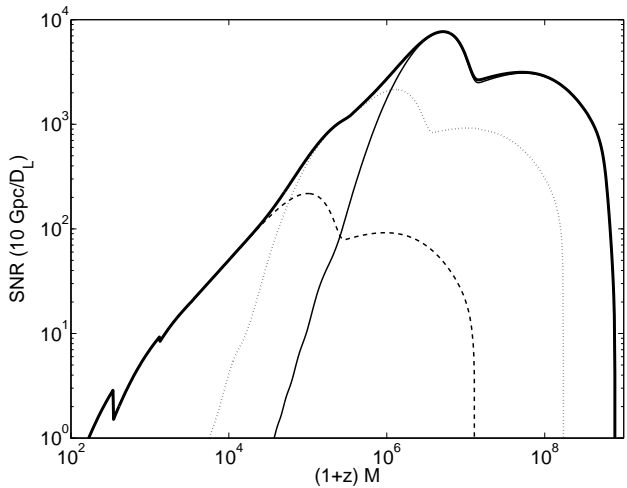}
\caption{SNR for sources at luminosity distance $D_L = 10$ Gpc plotted
  vs. redshifted mass for LISA.  The contributions from the early
  inspiral $-\infty < t < -1000M$ (dashed), late inspiral $-1000 < t <
  -50M$ (dotted), and merger-ringdown $-50M < t < \infty$ (thinner
  solid), as well as the SNR from the entire waveform (thicker solid)
  are shown.}
\label{fig:LISASNR}
\end{figure}

The SNR for LISA is shown as a function of redshifted mass, normalized
for $D_L=10{\rm Gpc}$ in Fig.~\ref{fig:LISASNR}.  The bump in the
curves is caused by the binary confusion noise.  Again we see the
enhancement of SNR from the merger-ringdown part (thin solid line) of
the waveforms, and confirm the strikingly large values of SNR
obtainable by LISA for these sources seen in \cite{Flanagan97a} and
\cite{Buonanno:2006ui}. For systems with redshifted mass
$(1+z)M<3\times10^4$, the early inspiral $t<-1000M$ portion of the
waveform dominates.  The highest SNRs for equal-mass nonspinning
mergers are obtained for systems with$(1+z)M>10^6$, again dominated by
the merger-ringdown portion of the waveform.
 
\begin{figure}
\includegraphics*[scale=.28, angle=0]{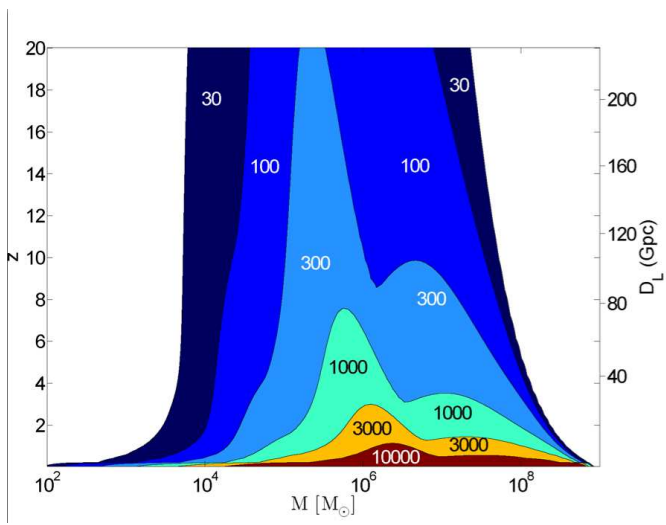}
\caption{SNR contour plot with mass and redshift dependence for LISA.
  Note that MBBHs with masses $M > 10^7 M_{\odot}$ may not coalesce
  within a Hubble time \cite{Milosavljevic:2002bn}.}
\label{fig:LISASNRcontour}
\end{figure}

Contours of SNR for LISA are shown in Fig.~\ref{fig:LISASNRcontour}
and demonstrate that LISA can observe MBBHs throughout the observable
universe at large SNRs.  We find it encouraging that, in addition to
the large SNR values predicted for LISA overall, some of the largest
values out to the largest redshifts occur in the mass range $10^5
M_{\odot} \leq M \leq 10^7 M_{\odot}$ where models of BBH populations
predict that the binaries can coalesce within a Hubble time
\cite{Milosavljevic:2002bn} and that the event rates for LISA are
several per year \cite{Sesana:2004gf}.  As discussed above, the
effects of unequal masses will tend to decrease these SNR values,
while spins may increase or decrease them.

\begin{figure}
\includegraphics*[scale=.35, angle=0]{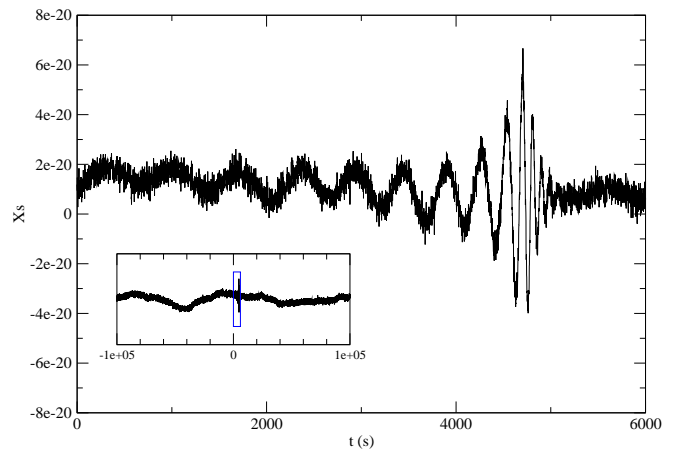}
\caption{Simulated LISA data stream showing LISA's response to a 
system of two equal-mass black holes ($M=10^5 M_{\odot}$) located at 
redshift z=15 observed on the system's equatorial plane.  The quantity
plotted is an unequal arm Michelson interferometer observable ``X'' \cite{Shaddock:2003dj}. 
The LISA response and instrumental noise are realized using the 
LISA Simulator \cite{Cornish:2002rt,LISASIM}, and colored noise was
added to represent the unresolvable galactic binary foreground with the
spectrum used in Ref.~\cite{Barack:2004}. The inset shows the signal over
a longer duration where low-frequency noise is evident. 
\label{fig:LISAsim}}
\end{figure}
Even for nonoptimal configurations, the presence of an MBBH
coalescence in the LISA data stream can dominate all the anticipated
noise sources.  Fig.~\ref{fig:LISAsim} shows a simulation of LISA's
response to the merger of equal-mass nonspinning black holes with
total mass $M=10^5 M_{\odot}$ located at redshift $z=15$, and oriented
so that LISA lies in the system's equatorial plane, where the
radiation is weakest. The SNR for a signal from such a source will be
$\sim 200$, averaged over sky positions and polarizations (see
Fig.~\ref{fig:LISASNRcontour}).

\section{Summary and Discussion}
\label{sec:summary}

Coalescing BBHs are expected to be the strongest sources for both
ground-based interferometers as well as the space-based LISA.  In
particular, the strong-field merger portion of the gravitational wave
signal produces an intense burst of radiation and has the highest
luminosity, emitting more energy per second than the combined
starlight emitted in the observable universe.

Recent breakthroughs in numerical relativity have opened a new era in
understanding the late stages of binary black hole coalescence.  We
now have a good understanding of the merger-ringdown signal, starting
$\sim 50M$ before the peak radiation amplitude, for equal-mass
nonspinning BBHs.  The late inspiral evolution, that is, more than a
few orbits prior to ringdown, is more challenging. Such simulations
require better numerical stability and more computational resources,
as well as higher accuracy to control the accumulated phase error.

In this paper, we presented new simulations of equal-mass nonspinning
BBHs starting in the late inspiral regime and covering approximately
an additional factor of three in frequency before the merger-ringdown.
We carried out runs at three resolutions, $h_f = 3M/64, 3M/80,$ and
$M/32$.  Our runs start with relatively low eccentricity and show good
convergence and conservation properties.  We have demonstrated the
stability and accuracy of our simulations over the course of an
unprecedented seven orbits.  We also showed the value of using
frequency (rather than time) to set a reference for the purpose of
comparing results between runs as well as with the PN approximation.
In recasting phase vs. frequency we have found particularly good
agreement, not only between the runs but also with PN predictions.

We have also matched the resulting gravitational waveforms to PN
calculations covering the earlier parts of the inspiral.  The
resulting full waveform has less than $3/4$ cycle of accumulated phase
error over its entire frequency band.  Using this waveform, we
calculated the SNRs for iLIGO, adLIGO, and LISA. Our results confirm
the importance of the merger-ringdown signal, which yields the highest
values of SNR for the majority of equal-mass signals
\cite{Flanagan97a,Buonanno:2006ui}.  We also show the SNR for the late
inspiral regime, which numerical simulations are now beginning to
address.  The late inspiral dominates the SNR for iLIGO and adLIGO for
the lower mass ($\lesssim$ a few $\times 10M_{\odot}$) stellar BBHs,
and the SNR for LISA generated by MBBHs with $M \sim 10^5 M_{\odot}$.
Contour plots of SNR as a function of $z$ and $M$ show that adLIGO can
achieve SNR $\gtrsim 10$ for IMBBHs out to nearly $z \sim 1$, and that
LISA can observe MBBHs at SNR $ > 100$ out to the earliest epochs of
structure formation at $z > 15$.

Our work has focused on equal-mass nonspinning BBHs.  Astrophysically,
BBHs are expected to have unequal masses and spins. In general, the
effects of unequal masses will tend to decrease the resulting SNRs,
while spins can increase them.  Calculations of merger-ringdown
waveforms for several mass ratios, and for some spins, are currently
available; we expect this to be a significant area of focus in the
foreseeable future, both expanding the range of parameters studied and
extending the duration of the resulting simulations.

As computational technologies mature, simulations of the merger can be
used in conjunction with gravitational wave observations to probe
gravity in the arena of strong fields.  In particular, if the binary
masses and spins can be obtained from observations of the inspiral (as
demonstrated, {\em e.g.}, in Ref. \cite{Lang:2006bz} for LISA),
numerical relativity can be used to calculate the merger
waveform. This will allow a comparison between the predictions of
general relativity -- or indeed, any other theory of gravity used in a
numerical simulation -- with observations in the regime of very strong
gravity.

\acknowledgments

We are pleased to acknowledge useful discussions with Chris
Belczynski, Scott Hughes, Cole Miller, Peter Shawhan, David Shoemaker,
and Tuck Stebbins.  This work was supported in part by NASA grant
O5-BEFS-05-0044.  The simulations were carried out using Project
Columbia at NASA Ames Research Center and the NASA Center for
Computational Sciences at Goddard Space Flight Center.  J.v.M., B.K.,
and M.K. were supported by the NASA Postdoctoral Program at the Oak
Ridge Associated Universities.  S.T.M. was supported in part by the
Leon A. Herreid Graduate Fellowship.  D.C. was supported in part by
The Korea Research Foundation and The Korean Federation of Science and
Technology Societies Grant funded by the Korean Government (MOEHRD,
Basic Research Promotion Fund).



\end{document}